\documentclass[11pt,english]{article}
\usepackage{rotating}
\usepackage{pstricks}
\usepackage{color}
\usepackage{axodraw}
\usepackage{latexsym}
\usepackage{graphicx}
\usepackage{psfrag}
\usepackage{amsmath,amssymb}
\makeatletter
\def\section{\@startsection {section}{1}{\z@}{-3.5ex plus -1ex minus
 -.2ex}{2.3ex plus .2ex}{\large\bf}}
\def\subsection{\@startsection{subsection}{2}{\z@}{-3.25ex plus -1ex
minus -.2ex}{1.5ex plus .2ex}{\normalsize\bf}}
\makeatother
\makeatletter

\@addtoreset{equation}{section}

\makeatother

\newcommand{\captionfonts}{\small}
\makeatletter  
\long\def\@makecaption#1#2{%
  \vskip\abovecaptionskip
  \sbox\@tempboxa{{\captionfonts #1: #2}}%
  \ifdim \wd\@tempboxa >\hsize
    {\captionfonts #1: #2\par}
  \else
    \hbox to\hsize{\hfil\box\@tempboxa\hfil}%
  \fi
  \vskip\belowcaptionskip}
\makeatother   
\def\Dslash{\hspace{3pt}\raisebox{1pt}{$\slash$} \hspace{-8pt} D}

\def\bea{\begin{eqnarray}} \def\eea{\end{eqnarray}}
\def\be{\begin{equation}} \def\ee{\end{equation}} \def\nn{\nonumber}
\def\a{& \hspace{-7pt}}  \def\Z{{\bf Z}}

\newcommand{\promille}{%
  \relax\ifmmode\promillezeichen
        \else\leavevmode\(\mathsurround=0pt\promillezeichen\)\fi}
\newcommand{\promillezeichen}{%
  \kern-.05em%
  \raise.5ex\hbox{\the\scriptfont0 0}%
  \kern-.15em/\kern-.15em%
  \lower.25ex\hbox{\the\scriptfont0 00}}

\begin{document}

\thispagestyle{empty}

\begin{center}
\hfill SISSA-19/2010/EP \\

\begin{center}

\vspace{1.7cm}

{\LARGE\bf  A Simple UV-Completion of QED in 5D}

\end{center}

\vspace{1.4cm}

{\bf  Roberto Iengo and Marco Serone}

\vspace{1.2cm}

{\em International School for Advanced Studies (SISSA) and Istituto Nazionale di Fisica Nucleare (INFN), Via Beirut 2-4, I-34151 Trieste, Italy} \\

\medskip

\end{center}

\vspace{0.8cm}

\centerline{\bf Abstract}
\vskip 10pt

We construct a Lifshitz-like version of five-dimensional (5D) QED which is UV - completed
and reduces at low energies to ordinary 5D QED. The UV quantum behaviour of this theory is very smooth. In particular, the gauge coupling constant is 
finite at all energy scales and at all orders in perturbation theory. We study the IR properties of this theory, when compactified on a circle, and
compare the one-loop energy dependence of the coupling in the Lifshitz theory with that
coming from the standard 5D QED effective field theory.  The range of validity of the 5D effective field theory is 
found to agree with the more conservative version of Naive Dimensional Analysis.

\vspace{2 mm}
\begin{quote}\small

\end{quote}

\vfill

\newpage

\section{Introduction}

Quantum field theories in more than four space-time dimensions have received a lot of attention in the past ten years. 
They allow us to address standard well-known problems in four-dimensional (4D) physics, such as the gauge and/or flavour hierarchy problems, from a different perspective, leading 
to novel scenarios, such as the possibility of having a fundamental TeV-sized quantum gravity scale \cite{ArkaniHamed:1998rs} or a TeV scale naturally generated by an extreme red-shift effect from a warped extra dimension \cite{Randall:1999ee}.  
Extra dimensional (ED) field theories are non-renormalizable and require an ultra-violet (UV) completion. 
There is little doubt that such UV-completions exist, in particular in string theory where ED are predicted and necessary. 
Constructing string theory models which reduce at low energies to the ED models considered in the literature is a difficult task.
As a matter of fact,  we do not know sufficiently simple and concrete UV completions of 
ED field theories. 

Aim of this paper is to concretely provide a UV completion of ED theories. For simplicity, we will focus our attention on a specific simple model which is 
QED in five dimensions (5D) compactified on a circle, although our construction is more general and can allow for a possible UV completion of any 5D (or higher) gauge theory. Our model is of the Lifshitz type \cite{Lif,Anselmi:2007ri}, where Lorentz invariance is explicitly broken at high energies. 
In these theories, the presence of higher derivative  (in the spatial directions only) quadratic terms improve the UV behavior of the particle propagator, without introducing ghost-like degrees of freedom. 
This kind of UV completion does not require the introduction of extra degrees of freedom, but rather modifies the propagation of the already existing ones.

The quantum UV behaviour of our Lifshitz 5D QED is incredibly simple. 
The photon anomalous dimension vanishes 
to all orders in perturbation theory and correspondingly the 
electric charge is completely finite.  This phenomenon is explained by the fact that the UV theory formally looks like a non-relativistic theory
for which particle anti-particle creation is suppressed at high energies.
Another marginal coupling in the theory, magnetic-like, is shown to be UV-free, so 
the theory is UV completed and perturbative at any energy scale (neglecting gravity, of course).

After having studied the UV one-loop renormalization of the theory, we turn our attention to its infrared (IR) behaviour.  
In particular, we show in some detail that the universal IR energy dependence of the finite gauge coupling (electric charge)
in the Lifshitz theory coincides with that computed in the effective field theory, as it should.
En passant, 
we explicitly show that the decoupling of heavy massive states in 5D effective theories is less efficient than in 4D; the effect of massive particles in 5D does not vanish as $1/M^2$,  like in 4D, but only as $1/M$, in agreement with the known fact that the sensitivity of an effective theory to its 
UV completion is higher in 5D than in 4D. 

The whole one-loop energy behaviour of the 4D inverse square coupling $g_4^{-2}$ is the following: 
starting from the IR, for energies much smaller than the compactification scale,
$g_4^{-2}$ decreases logarithmically, for a small window above the compactification scale it decreases linearly and then
for yet larger energies $g_4^{-2}$ tends to a constant, with the one-loop correction going to zero as 
$1/E^{1+\gamma}$, with $\gamma={\cal O}(g_4^2)>0$.

Estimates based on Naive Dimensional Analysis (NDA) \cite{NDA}  or on 
the unitarity of scattering amplitudes (see e.g. \cite{Chivukula:2003kq})  show that the energy range of validity of phenomenologically
interesting ED theories is quite limited, sometimes at the edge of not being present at all. 
By using our UV-completed model, we can better quantify the cut-off $\Lambda$ of the effective 5D QED, 
identifying it with the scale where the higher derivative Lifshitz operators become relevant. 
More precisely, $\Lambda$ is defined as the scale where  the UV-dependent one-loop photon vacuum polarization correction becomes 
of the same order as the calculable one in the effective 5D theory, with the asymptotic value of the coupling
still in the perturbative range.
The resulting cut-off turns out to be approximately equal to the one predicted by a conservative NDA estimate:
\be
\Lambda \gtrsim \frac{48\pi}{5}\frac{1}{R}\Big(\frac{1}{g_4^2}-\frac{1}{g_{4,\infty}^{2}}\Big)\,,
\label{LambdaLifInt}
\ee
with $g_4$ evaluated at the compactification scale $1/R$ and $g_{4,\infty}$ its asymptotic UV value, as computed in the Lifshitz theory. 
We are not  taking into account here many other effects that can sensitively
change the estimate of $\Lambda$, such as the number of particle species or, for $S^1/\Z_2$ orbifolds-interval compactifications, possible additional localized Lagrangian terms or warp factors.  When these effects are considered in phenomenologically promising 5D theories, the estimate
(\ref{LambdaLifInt}) is lowered by one order of magnitude or more. 

Lorentz invariance is explicitly and maximally broken in the UV in Lifshitz-type theories and it is not automatically
recovered in the IR, $E<\Lambda$, unless a fine tuning is imposed \cite{Iengo:2009ix,Collins:2004bp} or some dynamical mechanism advocated. 
Aside the above tuning, other experimental constraints, mainly of astrophysical origin, severely
constraint the parameters of our theory, which should not be considered too seriously as a realistic UV completion, at the present stage at least.
Yet we believe that the construction underlying it can be very useful to build renormalizable 5D theories.

The structure of the paper is as follows. In section 2, after a very brief review of the consruction of Lifshitz-like theories, we present our model.
In section 3 the Renormalization Group (RG) flow in the UV regime $E> \Lambda$ is studied and it is in particular shown that the theory is UV-completed and perturbative
at any energy scale in a wide range of parameter space. 
In section 4 we study the IR behaviour, $E<\Lambda$, of the gauge coupling in the Lifshitz theory.\footnote{A reader 
more interested to the IR properties of our model might want to skip sections 2 and 3 and go directly to section 4.}
We analyze the RG flow of the coupling from an effective 5D point of view and compare it with the UV model in subsections 4.1 and 4.2. In subsection 4.3 we determine the cut-off of the 5D effective theory as determined from the Lifshitz model.
In section 5  we show how Lorentz invariance can be recovered at low energies and briefly comment on astrophysical bounds.
In section 6 we conclude.  In appendix A an analytic formula
for the energy behaviour of the coupling is derived.

\section{The Model} 

The key point of the construction of Lifshitz-like theories is to break Lorentz invariance, so that one is allowed to introduce higher derivative
terms in the spatial derivatives and quadratic in the fields, without necessarily introducing
the dangerous higher time derivative terms  that would lead to violations  of unitarity.
The UV behavior of the propagator is improved and theories otherwise non-renormalizable
become effectively renormalizable. In Lifshitz-like theories an invariance under ``anisotropic"  scale transformations is imposed:
\be
t = \lambda^z t^\prime \,, \ \ \ \ \ x^i = \lambda x^{i\prime}\,, \ \ \ \ \
\phi(x^i,t) = \lambda^{\frac{z-d}{2}} \phi^\prime(x^{i\prime},t^\prime)\,,
\label{scalingW}
\ee
where $i=1,\ldots, d$ parametrizes the spatial directions, $\phi$ denotes a generic field and $z$ is an integer, sometimes called critical exponent. 
We will always assume to be in the preferred frame where spatial rotations and translations are unbroken symmetries.
According to eq.(\ref{scalingW}), we can assign to the coordinates and to the fields a ``weighted'' scaling dimension:
\be
[t]_w = - z\,, \ \ \ \  [x^i]_w = -1\,, \ \ \ \  [\phi]_w = \frac{d-z}{2}\,.
\label{weightedDim}
\ee
The renormalizability properties of Lifshitz-like theories have been extensively studied  for scalar, fermion \cite{Anselmi:2007ri} and gauge theories \cite{Anselmi2}.  
The usual power-counting argument for the renormalizability of a theory is essentially still valid, provided one substitutes the standard scaling dimensions of the operators  by their ``weighted scaling dimensions'' \cite{Anselmi:2007ri}, i.e. by the dimensions implied by the assignment (\ref{weightedDim}).

Here we will focus on QED in 5D.
The weighted dimensions of the photon and fermion fields are easily fixed by looking at their time component kinetic terms. We have
\be
[\psi]_w = \frac{d}{2}\,, \ \ \ \ \ \ [A_0]_w = \frac{d+z-2}{2}\, , \ \ \ \ \ [A_m]_w = \frac{d-z}{2}\,.
\label{scaleweighted}
\ee
Notice that $A_0$ and $A_m$ necessarily have different weighted dimensions, since no time derivative acts on $A_0$.
Equations (\ref{scaleweighted}) are consistent with gauge invariance, requiring  
\be
[g]_w = [\partial_0]_w - [A_0]_w=[\partial_m]_w - [A_m]_w= \frac{2-d+z}{2}\,.
\ee
Here and in what follows, $m,n,p$ are indices running over all spatial directions, including the compact directions.
The Lifshitz version of QED is hence renormalizable in $d$ spatial dimensions, provided that 
\be
z=d-2.
\ee 
In what follows we focus our attention on the case $z=2$ and $d=4$, i.e. 5D QED.
For simplicity, we take $\psi$ to be an  
ordinary Dirac field, despite the fact that at high energy the absence of $SO(5)$ invariance might allow to construct two independent
spinors related by $\gamma^0$, which is a sort of ``chirality" matrix.
In order to simplify the structure of the Lagrangian, we impose the conservation of a $\Z_2$ symmetry $C$, 
combination of the usual charge conjugation and parity in the extra direction:
\bea
\psi^C(x_0,x_i,y)  \a = \a C \bar \psi^T(x_0,x_i,-y)\,, \ \ \ 
A_0^C (x_0,x_i,y) = -A_0 (x_0,x_i,-y)\,, \nn  \\ 
A_i^C (x_0,x_i,y) \a= \a -A_i (x_0,x_i,-y)\,, \ \ \ \ A_y^C (x_0,x_i,y) =+ A_y (x_0,x_i,-y)\,,
\label{Z2}
\eea
where the matrix $C$ is the usual charge conjugation matrix for spinors as defined in $D=3+1$ Lorentz invariant theories, $y$ is the compact coordinate
and $i=1,2,3$ runs over the non-compact spatial directions.
The most general Lagrangian with weighted marginal and relevant couplings and invariant under (\ref{Z2}) is the following:
\be
{\cal L} =\frac 12 F_{m0}^2 - \frac{c_\gamma^2}{4} F_{mn}^2 - \frac{a_\gamma^2}{4\Lambda^2} (\partial_m F_{np})^2 + \bar \psi ( i \Dslash_0 -i c_\psi \Dslash -M) \psi
-\frac{a_\psi}{\Lambda} |D_m \psi|^2 -\frac{i\lambda}{\Lambda^{3/2}} F_{mn} \bar \psi \gamma^{mn} \psi\,,
\label{Lquad}
\ee
where $\Dslash_0 = \gamma^0 (\partial_0 - i (g/\sqrt{\Lambda}) A_0)$, $\Dslash = \gamma_m (\partial_m - i  (g/\sqrt{\Lambda})  A_m)$ and $\Lambda$ is a high-energy scale parametrizing the strength of the higher derivative operators. The symmetry (\ref{Z2}), in combination with $SO(d)$ rotational invariance, is crucial to get rid of several otherwise allowed terms, such as $\bar \psi \gamma^0 \psi$, $\bar \psi D_0 \psi$, $F_{mn}^3$, etc. Below the scale $\Lambda$ the theory defined by (\ref{Lquad}) flows to the usual 5D QED, provided that $c_\psi = c_\gamma$ to sufficient accuracy. For simplicity, in the following we will use units in which $\Lambda=1$.
\begin{figure}[t]
\begin{center}
\begin{picture}(400,320)(0,0)
\SetWidth{.8}
\ArrowLine(0,300)(100,300)
\Text(50,315)[]{\large{$p$}}
\Text(190,300)[]{\ \ $=\;${\Large{$\frac{i (\gamma^0 \omega - c_\psi \gamma^m p^m + a_\psi p^2 + M)}{\omega^2 - c_\psi^2 p^2 - (a_\psi p^2+M)^2} $}}}

\Text(50,235)[]{\large{$p$}}
\Text(10,235)[]{$0$}
\Text(90,235)[]{$0$}
\Photon(0,220)(100,220){3}{9}
\Text(132,220)[]{\ \ $=\,${\Large{$\frac{i}{p^2}$}}}

\Text(10,275)[]{$m$}
\Text(90,275)[]{$n$}
\Text(50,275)[]{\large{$p$}}
\Photon(0,260)(100,260){3}{9}
\Text(178,260)[]{\hspace{1.5cm}$=\Big(\delta_{mn}-\frac{p_m p_n}{p^2}\Big)${\Large{$\frac{i}{\omega^2-a_\gamma p^4-c_\gamma^2 p^2}$}}}

\Photon(0,100)(60,100){3}{5}
\ArrowLine(80,125)(60,100)
\ArrowLine(60,100)(80,75)
\Vertex(60,100){1.5}
\Text(224,100)[]{$=\,${{{$-i gc_\psi \gamma^m-i g a_\psi (p_2-p_1)^m-2i\lambda p_{3,n} \gamma^{mn}$}}}}
\Text(35,115)[]{$m$}
\Text(93,82)[]{$p_1$}
\Text(93,120)[]{$p_2$}
\Text(30,82)[]{$p_3$}

\Photon(0,165)(60,165){3}{5}
\ArrowLine(80,190)(60,165)
\ArrowLine(60,165)(80,140)
\Vertex(60,165){1.5}
\Text(135,165)[]{\ \ $=i g \gamma^0$}
\Text(35,180)[]{$0$}

\Photon(0,20)(100,20){3}{9}
\ArrowLine(20,50)(50,20)
\ArrowLine(50,20)(80,50)
\Vertex(50,20){1.5}
\Text(150,20)[]{{\ \ \ $=\,${$-2ig^2 a_\psi \delta_{mn}$}}}
\Text(25,10)[]{{$m$}}
\Text(75,10)[]{{$n$}}

\SetWidth{.4}
\end{picture}
\caption{Feynman rules of the Lifshitz 5D QED theory in the Coulomb gauge. The momenta are all incoming.}
\label{FeynRules}
\end{center}
\end{figure}

\section{UV Behaviour}

In this section we compute the one-loop evolution of the marginal couplings in the UV regime, $E\gg1$. 
We regularize the theory using dimensional regularization in the spatial directions only ($d=4-\epsilon$) and renormalize using a minimal subtraction scheme where only the poles in $1/\epsilon$ are subtracted, with no finite term. 
Being Lorentz symmetry absent in this regime, it is convenient to work in the Coulomb gauge 
\be
\partial_m A_m =0.
\ee
Compactification effects and all relevant couplings can be neglected at these scales, allowing us to easily perform
the integration over the virtual energy exchanged in the one-loop graphs. We then set  $c_\psi=c_\gamma=M=0$. 
The fermion propagator  can be written as 
\be
-i G_\psi^0(\omega,p)=\frac{P_+}{\omega-a_\psi  p^2+i\epsilon}+\frac{P_-}{-\omega-a_\psi  p^2+i \epsilon} \,,
\label{psiProp}
\ee
where $\omega$ and $p$ denote energy and momentum, respectively, and  $P_\pm = (1\pm\gamma^0)/2$. We have explicitly written the $i\epsilon$ factors for reasons that will soon be clear.
Similarly, the spatial components of the photon propagator are
\be
G_{\gamma,mn}^0(\omega, p) = i\Big(\delta_{mn}-\frac{p_m p_n}{p^2}\Big)\frac{1}{(\omega-a_\gamma p^2+i\epsilon)(\omega+a_\gamma p^2-i\epsilon)}\,.
\label{gammaProp}
\ee
In eqs.(\ref{psiProp}) and (\ref{gammaProp}) the superscript $0$ stands for $c_\psi=c_\gamma=0$.
As can be seen from the Feynman rules reported in  fig.\ref{FeynRules}, no $\omega$ terms appear in the vertices, so that the $\omega$ integration can easily be performed using the residue theorem.
By denoting $q$ and $\omega_q$ the virtual momentum and energy running in the loop diagram, the $\omega_q$ dependence will only appear through the
fermion and photon propagators.  All non-vanishing loop diagrams we will consider involve one photon propagator (whose momentum we identify with $q$) and one or two fermion propagators. In general, for $n$ fermion propagators, the loop graph will be proportional to
\be
K_n=\int\!\frac{d\omega_q}{2\pi} \prod_{i=1}^n\frac{ -iG_\psi^0(\omega_i+\alpha_i \omega_q, p_i+ \alpha_i q)}{(\omega_q-a_\gamma q^2+i\epsilon)(\omega_q+a_\gamma q^2-i\epsilon)}\,,
\label{Nfer1phot}
\ee
with $\omega_i$ and $p_i$ external energies and momenta and $\alpha_i$ some constants. 
Since $P_+ P_- = 0$, eq.(\ref{Nfer1phot}) reduces to two terms, proportional to $P_+$ and $P_-$.
The former (the latter) have all fermion poles in the lower-half  (upper-half) $\omega$-plane, so that by appropriately closing the contour
at infinity, we can always avoid all fermion poles. Hence we get
\be
K_n = \frac{i}{2a_\gamma q^2} \prod_{i=1}^n\left( \frac{P_+}{\alpha_i a_\gamma q^2 +a_\psi  (\alpha_i q+p_i)^2-\omega_i}+
 \frac{P_-}{\alpha_i a_\gamma q^2 +a_\psi  (\alpha_i q+p_i)^2+\omega_i}\right)\,.
 \label{Kn}
\ee

\subsection{Vacuum Polarization}

The vacuum polarization of the photon is easily computed in this theory and turns out to identically vanish!
The fermion loop given by two trilinear vertices  vanishes due to the integration over $\omega$,
since  we can always choose to close the contour of integration in the upper/lower half $\omega$--plane with no poles, as in eq.(\ref{Nfer1phot}).
The same result can be checked to hold in the euclidean. After standard manipulations, it is easy to see that the loop is proportional to
\be
\int^\infty_{-\infty} d\omega_E \frac{\omega_E (p_E+\omega_E) - a b}{(\omega_E^2+a^2)[(\omega_E+p_E)^2+b^2]} = 0,
\ee
vanishing for any positive $a$ and $b$. Here $p_E = i \omega$ and $\omega_E= i \omega_q$ are the Wick rotated energies. 
Interestingly enough, this result is not only valid at one-loop order but to all orders in perturbation theory.
Consider first higher loop graphs constructed from the one-loop graph by adding photon lines only. Since, for $c_\psi=0$, the trilinear and quartic vertices commute with $P_\pm$, any such graph with an arbitrary number of fermion propagators will split in the sum of two terms of the kind
\be
{\rm Tr} \prod_{i=1}^n V_i  \bigg[\frac{P_+}{\omega+\omega_\gamma(i)-a_\psi(p+k_\gamma(i))^2+i \epsilon}+\frac{P_-}{-\omega-\omega_\gamma(i)-a_\psi(p+k_\gamma(i))^2+i \epsilon}\bigg]\,,
\label{OmegazeroGen}
\ee
where $\omega$ and $p$ are the energy and momentum of the virtual fermion running in the loop, $\omega_\gamma(i)$ and $k_\gamma(i)$ are the sum of the energies and the momenta of the photons attached to the fermion lines 
at the vertices $V_j$ one encounters in arriving at the propagator $i$. 
Note that the UV-vertices $V_i$ (see fig.1) commute with $P_{\pm}$.
Exactly as before, eq.(\ref{OmegazeroGen}) identically vanishes when integrated over $\omega$. 

When the components of the external photon are spatial, there is in addition a tadpole graph associated with the quartic vertex (see fig.1).
This is trivially vanishing in dimensional regularization since (after Wick rotation) 
\be
\int d\omega d^d k \frac{a_\psi k^2+m}{\omega^2+ (a_\psi k^2+m)^2} \propto 
\int d^d k \;1=  0\,.
\label{Tadpole}
\ee
Since a single fermion loop, dressed with an arbitrary number of photons, vanishes, any graph with an arbitrary number of fermion loops clearly vanish as well.
These results also extend to the compact case, since the spatial momenta do not play any role in the argument, as long as $c_\psi=0$.

The apparently strange absence of any  quantum correction in the photon propagator when $c_\psi=0$  has a simple physical explanation in the total decoupling, in the Lifshitz regime, of particle and anti-particles. In other words, the decomposition of the propagator as in eq.(\ref{psiProp}) is telling us that electrons and positrons in the Lifshitz regime behave similarly to standard electrons and positrons in the non-relativistic low-energy regime. In particular, there is no way to create a particle-anti particle pair and hence conservation of the charge forbids any virtual pair production in the vacuum.

In conclusion, no radiative corrections at all (finite or infinite) are induced to the photon two-point function  when $c_\psi=0$.
It then follows that in the UV the $\beta$-functions associated with the coupling constant $g$ and the parameter $a_\gamma$, as well as the anomalous dimensions of $A_0$ and $A_m$, vanish to all orders 
in perturbation theory:
\be
\beta_g = \beta_{a_\gamma} = \gamma_{A_0} = \gamma_{A_m} = 0 \,.
\ee

\subsection{Fermion Propagator}

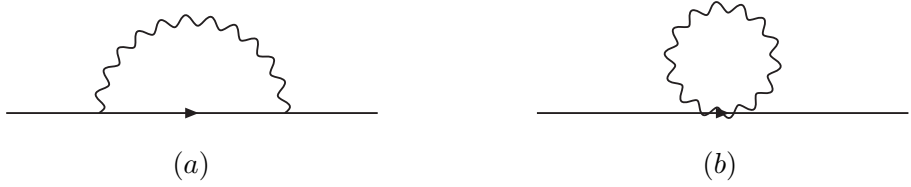
\begin{figure}[t]
\begin{center}
\begin{picture}(350,50)(0,0)
\SetWidth{.7}
   \SetColor{Black}
    \ArrowLine(0,20)(140,20)
    \PhotonArc(70,20)(35,0,180){2}{12}
\Text(70,0)[]{$(a)$}

    \ArrowLine(200,20)(340,20)
\Text(270,0)[]{$(b)$}
  \PhotonArc(270,40)(20,-90,270){2}{14}

\SetWidth{.4}
\end{picture}
\caption{One-loop graphs contributing to the correction of the fermion propagator.}
\label{1loopgraph}
\end{center}
\end{figure}

The one-loop correction to the fermion propagator $\Sigma(p)$ is given by the two graphs in fig.\ref{1loopgraph}. The tadpole graph (b), as well as the exchange of a virtual temporal photon in graph (a) are easily shown to vanish in dimensional regularization. The only non-trivial contribution is given by the exchange of spatial photons in graph (a). We get
\bea
\mbox{
  \begin{picture}(80,40) (6,17)
    \SetColor{Black}
    \ArrowLine(0,20)(80,20)
    \PhotonArc(40,20)(20,0,180){2}{9}
      \Text(30,10)[lb]{\footnotesize{\Black{$p+q$}}}
      \Text(0,10)[lb]{\footnotesize{\Black{$p$}}}
          \Text(5,40)[lb]{\footnotesize{\Black{$m,q$}}}
     \end{picture}} \a = \a\!\! \int\!\!\frac{d\omega_q}{2\pi}\frac{d^dq}{(2\pi)^d} G_{\gamma,ml}^0(\omega_q,q) G_\psi^0(\omega_p+\omega_q,p+q)
     \label{FP}  \\ \nn \\
     &\times &\!\! i\Big( g a_\psi (2 p+q)^m+2\lambda q_k \gamma^{mk}\Big) i\Big( g a_\psi (2 p+q)^l-2\lambda q_p \gamma^{lp}\Big) \,. \nn
 \eea
There is no need of introducing a Feynman parameter to compute the  graph (\ref{FP}). As previously explained, one can
first integrate over $\omega_q$,  using the residue theorem, and then over $q$ by using dimensional regularization.
Defining the counter-terms $\Delta Z_\psi$ and $\Delta Z_{a_\psi}$ as
\be
\mbox{
  \begin{picture}(10,20) (80,17)
    \SetColor{Black}
    \ArrowLine(0,20)(80,20)
\Vertex(50,20){3}
      \Text(40,10)[lb]{\footnotesize{\Black{$p$}}}
              \end{picture}}   =    i \gamma^0 \omega_p \Delta Z_\psi - i a_\psi p^2 \Delta Z_{a_\psi} \,,
  \label{Fct}
\ee
we get
\bea
\Delta Z_\psi & =  & -\frac{1}{\epsilon}\frac{12\lambda^2}{16 \pi^2 a_\gamma (a_\gamma+a_\psi)^2}\,, \nn \\
\Delta Z_{a_\psi} & = & \frac{1}{\epsilon} \frac{3}{16\pi^2}\Big(\frac{g^2 a_\psi}{a_\gamma(a_\gamma+a_\psi)} - \frac{4\lambda^2}{(a_\gamma+a_\psi)^3}\Big)\,,
\label{ZpsiZact}
\eea
from which one easily derives the $\beta$-function for $a_\psi$:
\be
\beta_{a_\psi}=\epsilon a_\psi (\Delta Z_{a_\psi}-\Delta Z_\psi)=\frac{3a_\psi^2}{16\pi^2} \bigg(\frac{g^2}{a_\gamma(a_\gamma+a_\psi)}+ \frac{4\lambda^2}{a_\gamma(a_\gamma+a_\psi)^3}\bigg)\,.
\label{betaapsi}
\ee

\subsection{The Electric Vertex}

In analogy to the path-integral derivation of the standard Lorentz invariant Ward Identity in scalar  QED,
one finds the following identity:
\be
\omega_p V^0(\omega_p,p,\omega_k,k) - p^m V^m(\omega_p,p,\omega_k,k)  = g \Big(G_\psi^f(\omega_p+\omega_k,p+k)^{-1} - G_\psi^f(\omega_k,k)^{-1} \Big) \,,
\label{wardEx}
\ee
where $V^0$ and $V^m$ are the full time and spatial components of the electric vertex, and $G^f_\psi$ is the full fermion propagator, including
all radiative corrections.  The one-loop corrections to the electric vertex are then related to the fermion counterterms in eq.(\ref{ZpsiZact}).
Denoting by  $\Delta V_0$ and $\Delta V^{m}$ the divergent terms of the one-loop corrections to the time and spatial components of the electric vertex, we have
\be
\Delta V_0= i g \gamma_0\Delta Z_{V_0} \,, ~~~~~ \Delta V^m=-i g a_\psi (2k+p)^m\Delta Z_{V_m}\,.
\ee
The Ward identity (\ref{wardEx}) immediately gives
\be 
\Delta Z_{V_0}=\Delta Z_\psi, ~~~~ \Delta Z_{V_m}=\Delta Z_{a_\psi}
\label{Deltaward}
\ee
Since there is no radiative correction to the photon propagator, the coupling constant $g$ is 
consistently not renormalized:
\be
g_0Z_\psi=gZ_{V_0} \rightarrow g_0=g\,,  ~~~~~~~~ g_0Z_{a_\psi}=gZ_{V_m} \rightarrow g_0=g \,.
\ee
We have checked, at one-loop level (for $c_\psi=c_\gamma=0$)  the validity of eqs.(\ref{wardEx})-(\ref{Deltaward}).

\subsection{The Magnetic Vertex}

The one-loop correction to the magnetic coupling $\lambda$ is given by the following graph:
\bea
\mbox{
  \begin{picture}(50,50) (10,35)
    \SetColor{Black}
\Photon(0,40)(30,40){2}{4}
\ArrowLine(55,70)(30,40)
\ArrowLine(30,40)(55,10)
\Photon(47,60.4)(47,19.6){2}{5}
\LongArrow(50,73)(35,55)
\LongArrow(5,31)(25,31)
\LongArrow(54,27)(54,52)
      \Text(60,40)[lb]{\footnotesize{\Black{$q$}}}
         \Text(35,65)[lb]{\footnotesize{\Black{$0$}}}
            \Text(20,48)[lb]{\footnotesize{\Black{$\lambda$}}}
      \Text(14,20)[lb]{\footnotesize{\Black{$m,p$}}}
            \end{picture}}  & = &  (-2i \lambda) \int\!\!\frac{d\omega_q}{2\pi}\frac{d^dq}{(2\pi)^d} G_{\gamma,nl}^0(\omega_q,q) G_\psi^0(\omega_p+\omega_q,p+q)
            G_\psi^0(\omega_q,q)
     \label{MagVertex}  \\ \nn \\
     &\times & i\Big(- g a_\psi q^n-2\lambda q_j \gamma^{nj}\Big)p_k \gamma^{mk} i\Big( -g a_\psi (2 p+q)^l+2\lambda q_i \gamma^{li}\Big) \,. \nn
 \eea
A straightforward computation fixes the counterterm to be
\bea
\mbox{
 \begin{picture}(20,50) (50,35)
    \SetColor{Black}
\Photon(0,40)(30,40){2}{4}
\ArrowLine(55,70)(30,40)
\Vertex(30,40){3}
\ArrowLine(30,40)(55,10)
\LongArrow(50,73)(35,55)
\LongArrow(5,31)(25,31)
         \Text(35,65)[lb]{\footnotesize{\Black{$0$}}}
      \Text(14,20)[lb]{\footnotesize{\Black{$m,p$}}}      
          \end{picture}}   & = &    -2i \lambda \Delta Z_\lambda p_k \gamma^{nk} \,, \ \ \ \ \Delta Z_\lambda = \frac{1}{\epsilon}\frac{4\lambda^2}{16\pi^2 a_\gamma (a_\gamma+a_\psi)^2} \,.
        \label{MagVertct}  \\        &&  \nn
 \eea
The $\lambda^2 g$ terms do not lead to divergences,  while those of the $\lambda g^2$ terms are exactly compensated by the fermion wave function counterterms 
in eq.(\ref{ZpsiZact}). We get
\be
\beta_\lambda =\epsilon \lambda (\Delta Z_\lambda-\Delta Z_\psi)= \frac{\lambda^3}{\pi^2 a_\gamma (a_\gamma+a_\psi)^2}\,.
\label{betalambda}
\ee

\subsection{RG  Flow in the Deep UV}
  
The UV evolution of couplings  is greatly simplified 
since $\beta_{a_\gamma} = \beta_g=0$. Only $a_\psi$ and $\lambda$ undergo a quantum 
evolution, according to eqs.(\ref{betaapsi}) and (\ref{betalambda}). At the perturbative level
no fixed points, other than the trivial $a_\psi = \lambda = 0$, may arise. For stability reasons, $a_\gamma$ must be positive, whereas
$a_\psi$ can have any sign.  For any choice of $a_\gamma$ and $a_\psi$, the magnetic coupling $\lambda$ grows in the UV, so that the theory
does not seem to be asymptotically free. However, the effective coupling constant in the theory, the one that controls the perturbative expansion,
is not $\lambda$, but a combination of $\lambda$ with $a_\gamma$ and $a_\psi$. Similarly for the gauge coupling constant $g$.
On dimensional grounds, the effective couplings are 
\be
\alpha \equiv  \frac{g^2}{a_\gamma} f_e\Big(\frac{a_\gamma}{a_\psi}\Big)\,, \ \ \ \ \ 
\beta \equiv \frac{\lambda^2}{a_\gamma a_\psi^2} f_m\Big(\frac{a_\gamma}{a_\psi}\Big)\,,
\label{alphabetadef}
\ee
where $f_e$ and $f_m$ are dimensionless functions, depending only on the ratio $a_\gamma/a_\psi$.
The theory is UV completed if, for $E\rightarrow\infty$, $\alpha$ and $\beta$ remain perturbative, and is UV free if
$\alpha,\beta\rightarrow 0$ in the limit.

Let us denote by $t=\log E/E_0$ and let us first assume that $a_\psi(0)>0$. In this case, $\beta_{a_\psi} >0$ and
$a_\psi$ grows in the UV. The parameter $a_\psi$ is not a proper coupling constant and perturbativity
is not necessarily lost when it grows large. It is lost only if $f_e$ and/or $f_m$ in eq.(\ref{alphabetadef}) grow with some powers of $a_\psi/a_\gamma$.
On the contrary, the regime of large $a_\psi$ is expected to be smoother and smoother, since $a_\psi$ enters also in the fermion propagators.
We now argue that $f_e$ and $f_m$ cannot have terms that grow like $a_\psi^n$, with $n>0$. 
Since no fermion loops are allowed, given any graph, any additional one-particle irreducible loop is obtained by adding a photon line and two vertices to a pre-existing fermion 
line,  thus obtaining two additional fermion propagators. The photon propagator gives no factors of $a_\psi$, while the two fermion propagators
a factor $1/a_\psi^2$. If both vertices are electric, proportional to $g a_\psi$, we then get $(g a_\psi)^2 \times 1/ a_\psi^2 = g^2$.
If they are both magnetic, we get  $\lambda^2 \times 1/a_\psi^2 = \lambda^2/a_\psi^2$. Hence for large $a_\psi$, $f_e$ and $f_m$ tend to a constant,
and the effective coupling constants of the theory become 
\be
\alpha_0 \equiv  \frac{g^2}{a_\gamma}\,, \ \ \ \ \ 
\beta_0 \equiv \frac{\lambda^2}{a_\gamma a_\psi^2}\,.
\ee
Being $a_\gamma$ and $g$ constant along the RG flow, 
\be
\alpha_0(t) = \alpha_0(0)
\ee
and $\alpha_0$, if small at $t=0$, remains so at all scales, without being asymptotically free.
Setting $a_\gamma=1$, for simplicity, the UV flow of $\beta_0$, for large $a_\psi$ is given by
\be
\dot{\beta_0} = \frac{1}{2\pi^2} \beta_0^2-\frac{3}{8\pi^2} g^2\beta_0\,,
\label{beta0}
\ee
where a dot stands for a derivative with respect to $t$. Solving eq.(\ref{beta0}), we get
\be
\beta_0(t) = e^{-\frac{3g^2 t}{8\pi^2}}  \beta_0(0) \bigg[1+\frac{4\beta_0(0)}{3g^2} (e^{-\frac{3g^2 t}{8\pi^2}}  -1)\bigg]^{-1}\,.
\label{beta0sol}
\ee
Equation (\ref{beta0sol}) shows that $\beta_0$ is asymptotically free in the UV, going to zero power-like, provided that
$4\beta_0(0)<3g^2$.

When $a_\psi(0)<0$, one has to distinguish the two cases $|a_\psi(0)|<a_\gamma$ and $|a_\psi(0)|> a_\gamma$. 
In the latter, $|a_\psi|$ grows and all the considerations made for $a_\psi>0$ apply. In particular, $\beta_0$ is UV free. 
In the former case, instead, $|a_\psi|$ decreases, $\beta_0$ increases and the coupling explodes in the UV.  

Summarizing, the theory is UV completed and perturbative for $4\beta_0(0)<3g^2$,
$a_\psi>0$ and $a_\psi< -a_\gamma$. The effective magnetic coupling $\beta_0$ goes to zero, while the electric one $\alpha_0$ remains constant. 

\subsection{Relevant Couplings}

In this subsection, for completeness,  
we compute the UV evolution of the parameters $c_\psi$ and $c_\gamma$ in eq.(\ref{Lquad}).
This is easily done by considering again the one-loop fermion and photon corrections, when these parameters are non-vanishing. For simplicity, we will set the magnetic coupling and the fermion mass to zero, $\lambda = M=0$.

Let us first consider $c_\psi$. The counterterm (\ref{Fct}) contains an extra term, so that now
\be
\mbox{
  \begin{picture}(10,20) (80,17)
    \SetColor{Black}
    \ArrowLine(0,20)(80,20)
\Vertex(50,20){3}
      \Text(40,10)[lb]{\footnotesize{\Black{$p$}}}
              \end{picture}}   =    i \gamma^0 \omega_p \Delta Z_\psi - i c_\psi p_m \gamma_m \Delta Z_{c_\psi} - i a_\psi p^2 \Delta Z_{a_\psi} \,.
\label{Fct2}
\ee
After a straightforward computation, we get
\be
\Delta Z_{c_\psi} = \frac{1}{\epsilon} \frac{g^2}{32\pi^2} \frac{a_\psi a_\gamma+a_\gamma^2-6 a_\psi^2}{a_\psi a_\gamma (a_\psi+a_\gamma)}\,.
\label{DeltaZc}
\ee
Since $\Delta Z_\psi=0$ when $\lambda=0$, eq.(\ref{DeltaZc}) directly gives the $\beta$-function for $c_\psi$:
\be
\beta_{c_\psi} = \epsilon c_\psi \Delta Z_{c_\psi}= \frac{g^2 c_\psi}{32\pi^2} \frac{a_\psi a_\gamma+a_\gamma^2-6 a_\psi^2}{a_\psi a_\gamma (a_\psi+a_\gamma)} \,.
\label{betacpsiUV}
\ee

The RG evolution of $c_\gamma$ is derived by computing the vacuum polarization of the spatial photon components  $\Pi_{mn}=\Pi_{mn}^1+\Pi_{mn}^2$:
\bea
\mbox{
  \begin{picture}(90,50) (10,37)
    \SetColor{Black}
\Photon(0,40)(30,40){2}{4}
\ArrowArc(50,40)(20,0,180)
\ArrowArc(50,40)(20,180,360)
\Photon(70,40)(100,40){2}{4}
           \Text(40,65)[lb]{\footnotesize{\Black{$q$}}}
                   \Text(35,5)[lb]{\footnotesize{\Black{$p+q$}}}
            \Text(8,48)[lb]{\footnotesize{\Black{$m,p$}}}
      \Text(76,48)[lb]{\footnotesize{\Black{$n,p$}}}
            \end{picture}}  & \equiv & \Pi_{mn}^1 =   (-1)g^2\!\! \int\!\!\frac{d\omega_q}{2\pi}\frac{d^dq}{(2\pi)^d}{\rm Tr}\bigg[ i\Big(c_\psi \gamma^m+a_\psi(p+2q)^m\Big) 
             \label{MagVertex} \nn  \\ \nn \\
     &\times & \!\!\! G_\psi(\omega_p+\omega_q,p+q)   i\Big(c_\psi \gamma^n+a_\psi(p+2q)^n\Big)  G_\psi(\omega_q,q) \bigg] \,. 
\eea
We do not write the tadpole contribution  $\Pi^2_{mn}$ to $\Pi_{mn}$, given by the contraction of the fermion lines in the quartic interaction (see fig.1).
It is energy and momentum independent and it is easily shown to ensure that $\Pi_{mn}(0,0)$ vanishes, as dictated by gauge invariance.
Spatial $SO(4)$ rotations and time reversal allow to write $\Pi_{mn}$ in the following form:
\be
\Pi_{mn}(\omega,p) =i \delta_{mn} \omega^2 f_1(\omega^2,p^2) + i(p_m p_n - \delta_{mn} p^2) f_2(\omega^2,p^2)\,,
\label{f12Def}
\ee
with $f_1$ and $f_2$ real functions. As already discussed, when $c_\psi = 0$, all photon vacuum polarization terms vanish. Hence $f_1$ and $f_2$ must be proportional to $c_\psi^2$.  
On dimensional grounds,\footnote{Recall that in the Lifshitz regime $c_\psi$ and $p$ have effectively the dimension of a mass, while $\omega$ of a mass squared.} this implies that the function $f_1$ is finite. On the other hand, the function $f_2$, associated with the operator $F_{mn}^2$, can develop a logarithmic divergence, 
which will be responsible for the RG evolution of $c_\gamma^2$. We leave to the next section a detailed study of the function $f_1$ and focus now on the
computation of the divergence of $f_2$. The latter is easily computed by taking the $c_\psi^2$ term of $f_2(0,0)$. 
The counterterm $\Delta Z_{c_\gamma}$ cancelling the above divergence is found
\be
\mbox{
  \begin{picture}(10,20) (80,17)
    \SetColor{Black}
    \Photon(0,20)(80,20){2}{10}
\Vertex(40,20){3}
      \Text(10,26)[lb]{\footnotesize{\Black{$m$}}}
   \Text(66,26)[lb]{\footnotesize{\Black{$n$}}}
      \Text(38,6)[lb]{\footnotesize{\Black{$p$}}}
              \end{picture}}   =    i(p_m p_n - \delta_{mn} p^2) c_\gamma^2 \Delta Z_{c_\gamma} \,, \ \ \ \   \Delta Z_{c_\gamma} = 
              - \frac{1}{\epsilon}\frac{g^2 c_\psi^2}{8\pi^2 a_\psi c_\gamma^2}\,,
 \label{FctGamma}
\ee
from which we get
\be
\beta_{c_\gamma^2} = \epsilon c_\gamma^2 \Delta Z_{c_\gamma} = - \frac{g^2 c_\psi^2}{8\pi^2 a_\psi}\,.
\label{betacgammaUV}
\ee
The RG eqs.(\ref{betaapsi}), (\ref{betacpsiUV}) and (\ref{betacgammaUV}) can easily be solved in the deep UV regime when
$a_\psi\gg a_\gamma$, in which case we approximately have
\bea
a_\psi(t) \a\simeq\a a_\psi(0)\, e^{\frac{3g^2}{16\pi^2}t}\,, \nn \\
c_\psi^2(t) \a \simeq \a c_\psi^2(0) \, e^{-\frac{3g^2}{8\pi^2}t} \,, \label{cevolution} \\
c_\gamma^2(t) \a\simeq \a c_\gamma^2(0) +\frac{2}{9} \frac{c_\psi^2(0)}{a_\psi(0)} \Big(e^{-\frac{9g^2}{16\pi^2}t}-1\Big) \,. \nn
\eea
Notice that the UV RG behaviour of $c_\psi^2$ and $c_\gamma^2$ does not significantly affect the physical speed $v=d\omega/dp$
of the photon and fermion, that at these energies is dominated by the classical Lifshitz-regime and linearly increases with energy, $v\simeq 2 \omega$.

\section{IR Behaviour and Connection with 5D Effective Theories}

We now consider the behaviour of the Lifshitz theory for  $E\ll \Lambda$. 
In particular, we will focus our attention on the energy behaviour of the finite function $f_1$, defined in eq.(\ref{f12Def}).
When the mass of the 5D fermion vanishes, two different regimes arise in the IR: i) $1/R \ll E \ll1$, where the theory is reliably described by its uncompactified 5D version and
ii) $E \leq 1/R \ll 1$, where compactification effects cannot be neglected. In order to distinguish the two regimes, we will denote the latter as deep IR regime.
In this section we will distinguish 5D and 4D gauge couplings by a subscript. What we have so far denoted by $g$ is replaced by $g_5$
and the 4D coupling is defined as
\be
g_5^2(E) \equiv 2\pi R \, g_4^{2}(E)\,.
\label{g45dDef}
\ee

\subsection{RG Behaviour of the Coupling in 5D Effective Field Theories}

The RG evolution of the inverse square gauge coupling in usual Lorentz-invariant  5D models
is expected to have the usual logarithmic behaviour for $E<1/R$, when the theory reduces to a 4D gauge theory for the zero modes, and a linear energy dependence for $E>1/R$.
Due to the presence of the infinite tower of massive Kaluza-Klein (KK) states,  with increasing mass, schemes such as momentum subtraction are preferred to schemes such as minimal subtraction,
in the regime $E>1/R$.\footnote{On the other hand, there are no problems in using dimensional regularization
if one is interested in the evolution of the coupling below $1/R$. KK modes can be integrated out and  their
 threshold corrections reliably computed in dimensional regularization, see e.g. \cite{Contino:2001si}.}
The contribution due to of a massive 4D fermion to the mass-dependent $\beta$-function of the 4D gauge coupling $g_4$ is well-known (see i.e. \cite{Weinberg2}):
\be
\beta_n = \frac{g_4^3}{2\pi^2}\int_0^1 \! dx \frac{x^2(1-x)^2 E^2}{m_n^2+E^2 x(1-x)}\,,
\ee
where $E$ is the sliding RG (euclidean) energy scale and $m_n^2=M^2+n^2/R^2$ is the mass of the KK mode $n$, where we keep for the moment the 5D bulk mass $M$. For (anti-)periodic fermions, $n$ is an (half-)integer. The total $\beta$-function is simply given by summing over all possible KK modes: $\beta_g=\sum_{n=-\infty}^\infty\beta_n$. 
Performing the sum, we get
\be
\beta_g = \frac{g_4^3R}{2\pi} \int_0^1 \!dx \frac{x^2(1-x)^2 E^2}{\sqrt{M^2+E^2 x(1-x)}}
\left\{
\begin{array}{l}
\coth \pi R \sqrt{M^2+E^2 x(1-x)} \,, \ \ n\in Z, \\
\tanh \pi R \sqrt{M^2+E^2 x(1-x)} \,,  \ \ n\in Z+\frac 12. \\
\end{array}
\right.
\label{betaeffective}
\ee
Eq.(\ref{betaeffective}) gives the following RG behaviour for $g_4^{-2}$: 
\be
g_4^{-2}(E) = g_4^{-2}(E_0) -\frac{1}{\pi^2} \int_0^1 \!dx x(1-x)\left\{ \begin{array}{l}
\log\frac{\sinh( \pi R \sqrt{E^2 x(1-x)+M^2})}{\sinh( \pi R \sqrt{E_0^2 x(1-x)+M^2})} \,, \ \ \ n\in Z, \\
\log\frac{\cosh( \pi R \sqrt{E^2x(1-x)+M^2})}{\cosh(\pi R \sqrt{E_0^2x(1-x)+M^2})} \,,  \ \ \ n\in Z+\frac 12. \\
\end{array}
\right.
\label{RGeffective}
\ee
It is useful to see in detail the regimes of vanishing compactification radius, $R\rightarrow 0$, and of decompactification limit, $R\rightarrow\infty$, for
$M=0$ and $M\gg E$.
When $R\rightarrow 0$, independently of $M$, for antiperiodic fermions, the argument of the logarithmic term approaches one,
giving no running at all, as expected, being all KK modes decoupled in this limit. On the other hand, for periodic fermions, we get
\bea
g_4^{-2}(E)\a=\a g_4^{-2}(E_0) -\frac{1}{6\pi^2}\log \frac{E}{E_0}\,, \ \ R\rightarrow 0\,, \ \ M = 0, \label{RGg4d}  \\
g_4^{-2}(E)\a=\a g_4^{-2}(E_0) -\frac{1}{60\pi^2}\frac{E^2}{M^2}+{\cal O}\Big( \frac{E^4}{M^4}\Big)\,, \ \ R\rightarrow 0\,, \ \  M \gg E  \gg E_0,
\label{RGg4dMinf}
\eea
which reproduce the usual logarithmic contribution due to a massless zero mode and the usual $1/M^2$ decoupling of a massive state in 4D.
When $R\rightarrow \infty$, the hyperbolic trigonometric functions in the argument of the logarithmic term become equal and for both periodic and anti-periodic fermions we get, for $M=0$,
\be
g_5^{-2}(E)= g_5^{-2}(E_0) -\frac{3}{256\pi} (E-E_0)\,, \ \ R\rightarrow \infty\,, \ \ M=0,  \label{RGg5d} 
\ee
where we have used eq.(\ref{g45dDef}). It is important to notice that  eq.(\ref{RGg5d}) does not give a quantitatively trustable behaviour of the coupling constant beyond one-loop level. Indeed, at two-loop level, by dimensional analysis and unitarity, a contribution to the r.h.s of eq.(\ref{RGg5d}) $\sim g^2_5(E_0) E^2 \log E R$ is expected. If we expand in $g^2_5(E_0)$, the two-loop term would be comparable or larger than the second iteration of the 1-loop term.
The RG flow for $1/R<E<1$ is hence not very useful.\footnote{We thank R. Rattazzi for this observation.} This issue is however not  important for what follows, because in our UV completion of the theory
at arbitrarily high energies we demand to always remain in the perturbative regime.

The uncompactified and large mass limit is taken by demanding
$MR\gg ER\gg E_0R$. In this limit we get
\be
g_4^{-2}(E)= g_4^{-2}(E_0)  - \frac{ER}{60 \pi}\frac{E}{M}+{\cal O}\Big(\frac{E^4 R}{M^3}\Big)\,, \ \ MR\gg E R\gg E_0 R\,.
\label{RGg5dMinf}
\ee
The decoupling in 5D is not as efficient as in 4D, with the heavy particle
effects vanishing as $1/M$, as opposed to $1/M^2$, as in eq.(\ref{RGg4dMinf}). In the 5D regime with $E\gg 1/R$, the factor of $R$ in the numerator
of eq.(\ref{RGg5dMinf}) enhances the effect.  Contrary to usual 4D theories, massive particles in 5D start to give a significant contribution to the gauge coupling
evolution at energies  well before their masses.\footnote{An equation very similar to eq.(\ref{RGeffective}) has been computed in \cite{Goldberger:2002cz} in dimensional regularization, but the $1/M$ decoupling of heavy states was not emphasized, being decoupling not manifest in that scheme.}

\subsection{One-loop Coupling Behaviour in the Lifshitz Theory}

The energy behaviour of the coupling in the Lifshitz model is obtained by studying the function $f_1$, defined in eq.(\ref{f12Def}).
Gauge invariance implies the following relations among the different photon polarization terms: 
\bea
&& \omega \Pi_{00} - p_m \Pi_{m0} = 0, \nn \\
&&  \omega \Pi_{0n}-p_m \Pi_{mn} = 0.
\label{PiPhoton}
\eea
In the parametrization (\ref{f12Def}) of $\Pi_{mn}$,  eq.(\ref{PiPhoton}) fixes
\be
\Pi_{00} = i p^2 f_1(\omega^2,p^2)\,, \ \ \ \ \  \Pi_{0m} =  i \omega p_m f_1(\omega^2,p^2)\,.
\label{Pi00}
\ee
The function $f_1$ is responsible for a finite renormalization of the photon kinetic term. The redefinition of the photon field
$A\rightarrow A/\sqrt{1+f_1} $ brings the kinetic term back in canonical form, but in so doing we get a rescaled coupling constant
\be
g_5^2(\omega,p^2) = \frac{g_5^2(\omega_0,p_0^2)}{1+f_1(\omega^2,p^2)-f_1(\omega_0^2,p_0^2)}\,,
\label{g5resum}
\ee
where $\omega_0$ and $p_0$ are arbitrary. Due to the Ward identity (\ref{wardEx}) there is no further renormalization of the coupling.
In the IR, by fine-tuning $c_\psi = c_\gamma$ with sufficient accuracy (see subsection 4.4 below), our model
flows to a Lorentz invariant 5D theory. It is then unnecessary to study $g_5^2$ as a function of both energy and momentum. We set, after having extracted the appropriate powers of $p$ as given by eq.(\ref{Pi00}),  $ p=0$ and study the gauge coupling as a function of the energy only. We take vanishing momentum of the photon in the compact space. 
The function $f_1$ is most easily computed by looking at the $\Pi_{00}$ component
of the photon vacuum polarization for euclidean energies $\omega = iE$. We have
\bea
\mbox{
  \begin{picture}(84,40) (10,37)
    \SetColor{Black}
\Photon(0,40)(30,40){2}{4}
\ArrowArc(46,40)(16,0,180)
\ArrowArc(46,40)(16,180,360)
\Photon(62,40)(92,40){2}{4}
           \Text(44,65)[lb]{\footnotesize{\Black{$q$}}}
                   \Text(35,5)[lb]{\footnotesize{\Black{$p+q$}}}
            \Text(13,48)[lb]{\footnotesize{\Black{$0$}}}
      \Text(80,48)[lb]{\footnotesize{\Black{$0$}}}
            \end{picture}}  \a \equiv \a \Pi_{00} =   (-1)g_4^2\!\! \sum_{n=-\infty}^\infty\int\!\!\frac{d\omega_q}{2\pi}\frac{d^3q}{(2\pi)^3}{\rm Tr}\bigg[ i \gamma^0
             \label{MagVertex} G_\psi(i E+\omega_q,p+q)   i \gamma^0  G_\psi(\omega_q,q) \bigg] \nn \\ \nn \\ \nn\\
             && \hspace{-3.8cm}=-4ig_4^2\!\!\sum_{n=-\infty}^\infty  \int\!\!\frac{d\omega_E}{2\pi}\frac{d^3q}{(2\pi)^3}\int_0^1\!\!dx\, \frac{-\omega^2_E +E^2 x(1-x) + a_\psi^2 q^2 (p+q)^2 +c_\psi^2 (q^2+p\cdot q)}{\big[\omega_E^2+E^2 x(1-x)+r(p,q)\big]^2} \,,
 \label{Pi00d}
\eea
with 
\be
r(p,q) = a_\psi^2 q^4 (1-x) + a_\psi^2 (p+q)^4 x +c_\psi^2 q^2 (1-x) +c_\psi^2 (p+q)^2 x\,, \ \ \ q^2 = q_i q_i+\frac{n^2}{R^2}\,.
\ee
\begin{figure}[t!]
\begin{minipage}[t]{0.465\linewidth} 
\begin{center}
\includegraphics*[width=\textwidth]{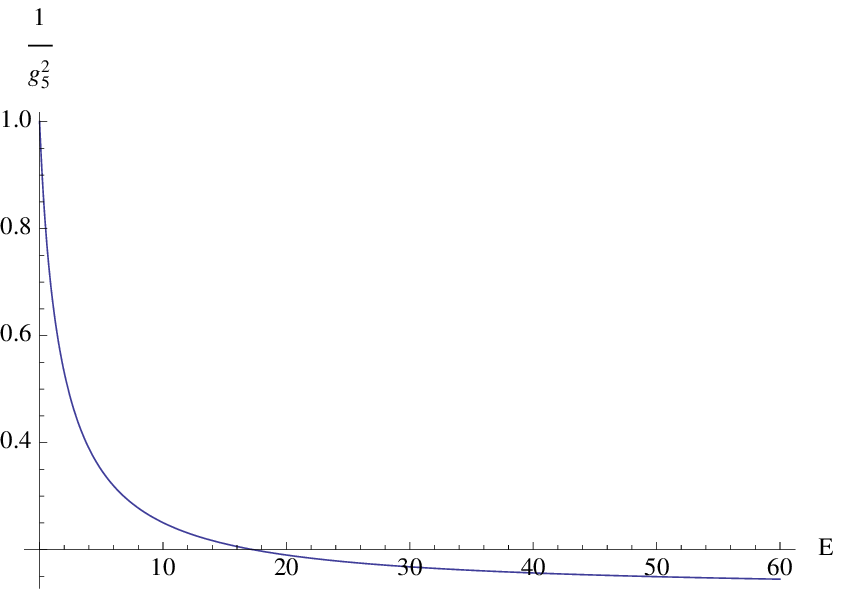}
\end{center}
\end{minipage}
\hspace{0.5cm} 
\begin{minipage}[t]{0.48\linewidth}
\begin{center}
\includegraphics*[width=\textwidth]{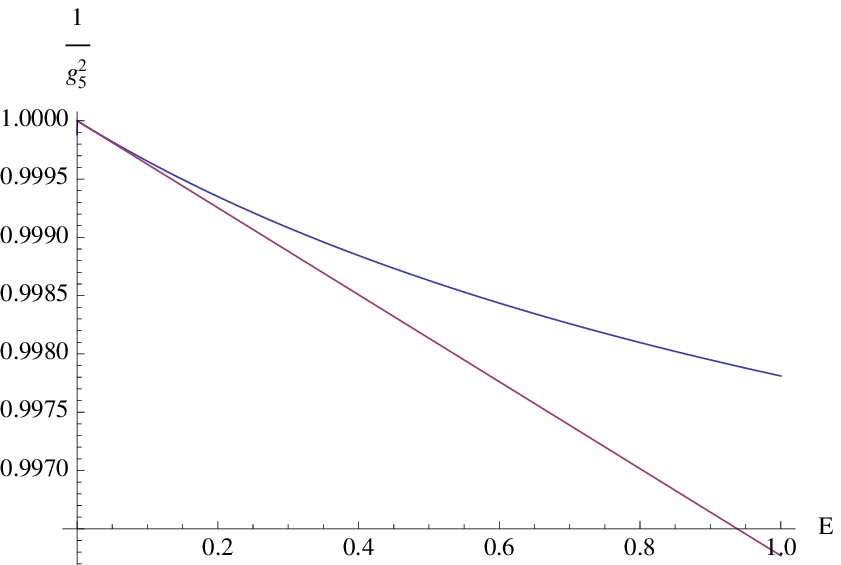}
\end{center}
\end{minipage}
\caption{Left panel: $g_5^{-2}$ as a function of the energy scale $E$ in the uncompactified Lifshitz Theory. Right panel: Comparison between the Lifshitz behaviour (blu line) with the 5D linear regime (\ref{RGg5d}) (red straight line). The energy is in units of $\Lambda$. We have taken $E_0=0$,  $g_5^{-2}(0)=1$ and $a_\psi=1$.} 
\label{figLifg}
\end{figure}
Expanding up to quadratic terms in the spatial momentum $p$, and by integrating over $\omega_E$ and afterwars over $x$, we  get the desired form of the function $f_1(iE,0)$:
\be
f_4(E,1/R)\equiv g_4^{-2}f_1(iE,0) = \frac{1}{3 \pi^2 c_\psi} \sum_{n=-\infty}^\infty \int_0^\infty \!\! \frac{s^2(3\tilde s^4+3\tilde s^2-s^2)ds}{\tilde s^3 (1+\tilde s^2)^{3/2}(4\tilde s^2+4\tilde s^4 +\mu^2)}\,,
\label{f4D}
\ee
with 
\be
\mu = \frac{a_\psi E}{c_\psi^2} \,, \ \ \ \ \ \  s= \frac{q a_\psi}{c_\psi}\,, \ \ \ \ \ 
\tilde s^2 = s^2 + \frac{a_\psi^2 n^2}{c_\psi^2 R^2}\,.
\label{musDef}
\ee 
At  {\it any} energy scale, $f_4(E,1/R)$ gives the one-loop energy behaviour
of the gauge coupling:
\be
g_4^{-2}(E) = g_4^{-2}(E_0) + f_4(E,1/R) - f_4(E_0,1/R)\,.
\label{g25dE}
\ee
When $E\gg 1/R$, the sum over the KK  modes in eq.(\ref{f4D}) is reliably approximated by a continuos integration and we get
\be
f_{5}(E)\equiv g_5^{-2} f_1(iE,0) \simeq \frac{1}{16 \pi^2 a_\psi}\int_0^\infty \!\! \frac{s^2(3+4s^2)ds}{(1+s^2)^{3/2}(4s^2+4s^4 +\mu^2)}\,, \ \ \ E\gg 1/R\,.
\label{f5D}
\ee
Independently of the UV-completion of the theory, the energy dependence
of the coupling constant at low energies should be dictated by its RG evolution.
Indeed, the logarithmic and linear terms in eqs.(\ref{RGg4d}) and (\ref{RGg5d}) are non-analytic in $E^2$, due to the IR behavior of the integrand,
and thus calculable.  We did not find an explicit formula for $f_4(E,1/R)$, but its deep IR
behaviour  can be extracted by considering the $n=0$ $R$-independent KK mode only 
and noticing that the IR divergence of the  integrand comes from
small values of $s$ so that we can simplify the integrand. We get
\begin{figure}[t!]
\begin{center}
\includegraphics*[width=0.6\textwidth]{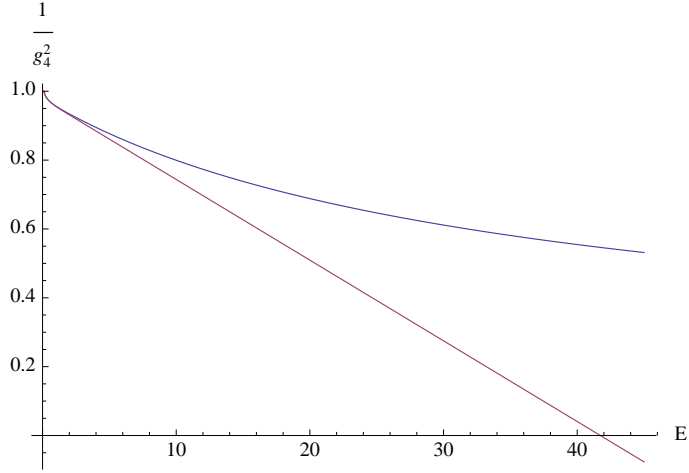}
\end{center}
\caption{Comparison between the exact Lifshitz behaviour as given by eq.(\ref{f4D}) (blu line) with the effective coupling (\ref{RGeffective})  for periodic fermions with $M=0$ (red line). 
The energy is in units of $1/R$. We have taken $E_0=1/(10R)$,  $g_4^{-2}(1/(10 R))=1$, $a_\psi=1$ and $\Lambda = 25/R$.} 
\label{figLifgNum}
\end{figure}
\be
f_4(E) -f_4(E_0) \simeq  \frac{2}{3 \pi^2c_\psi} \int_0^\infty \!\! \frac{s(\mu_0^2-\mu^2)ds}{(4 s^2 +\mu^2)(4s^2+\mu_0^2)}=-\frac{1}{6\pi^2c_\psi}\log\frac{E}{E_0}\,, \ \ E,E_0\ll 
\frac{c_\psi}{R}\,,
\label{fEdeepIR}
\ee
reproducing eq.(\ref{RGg4d}) for $c_\psi=1$. In the uncompactified limit, it is possible to write an approximate analytic formula for $f_{5}(E)$ (which gives the exact limit in both the UV and the IR)
 from which the IR behaviour (\ref{RGg5d}) is computed:
\be
f_{5}(E) \simeq \frac{1}{16\pi^2 a_\psi} \bigg[\int_0^1\!ds \frac{3s^2}{(4s^2 +\mu^2)}+\int_1^\infty\!ds \frac{4s}{(4s^4 +\mu^2)}\bigg]=
\frac{4\pi+6\mu-(8+3\mu^2)\arctan\Big(\frac{2}{\mu}\Big)}{128\pi^2 a_\psi\mu} \,.
\label{roughf5d}
\ee
Expanding eq.(\ref{roughf5d})  for small $\mu$, one finds
\be
f_{5}(E)-f_{5}(E_0) = -\frac{3}{256\pi c_\psi^2}(E-E_0)+{\cal O}(E^2)\,, \ \ \ E\ll \frac{c_\psi}{a_\psi},
\ee
reproducing the coefficient in eq.(\ref{RGg5d}), with $c_\psi=1$.

In the deep UV regime, $E\gg 1$, eq.(\ref{roughf5d}) gives
\be
 f_5(E) =\frac{c_\psi^2}{32\pi a_\psi^2}\frac{1}{E}++{\cal O}\Big(\frac{1}{E^2}\Big)\,,
\label{f5EUV}
\ee
showing that the correction to the coupling vanishes like $1/E$ for $E\rightarrow \infty$.\footnote{The correction actually vanishes as $1/E^{1+3g^2/(4\pi^2)}$, due to the UV RG evolution of $c_\psi$ and $a_\psi$, as given by eq.(\ref{cevolution}).}

At a more quantitative level, eq.(\ref{roughf5d}) is not a very accurate approximation of $f_{5}(E)$ in the whole $E$ range. A more reliable, but more complicated,
analytic expression is reported in  appendix, see eq.(\ref{AnafE}).
We plot, for illustration, $g_5^{-2}(E)$  in the 5D uncompactified limit, as given by eqs.(\ref{RGg5d}) and (\ref{f5D}) (fig. 3)
 and $g_4^{-2}(E)$ in the 4D compact case, as given by eqs.(\ref{RGeffective}) and (\ref{f4D}) (fig. 4).

\subsection{Cut-off of the Effective Theory and Comparison with NDA Estimates}

The range of validity of 5D theories as perturbative and calculable effective field theories
is typically estimated, in absence of a concrete UV completion, by using Naive Dimensional Analysis \cite{NDA}. 
A possible definition of the maximum energy scale $\Lambda$ above which the theory breaks down
is derived from the photon vacuum polarization term.\footnote{Strictly speaking, one should use physical observables to identify the breakdown
of the theory, yet we believe that the photon propagator is a reliable quantity to look at.}
 When the one-loop correction becomes of the same order as the tree-level term,
calculability is certainly lost. A naive often used estimate takes just into account the phase space of the loop integration, taken as in 5D uncompactified space,
giving 
\be
\frac{g_5^2 \Lambda^{(1)}}{24\pi^3} = 1 \ \ \ \Longrightarrow \Lambda^{(1)} = \frac{12\pi^2}{g_4^2} \frac{1}{R}\,,
\label{NDA1}
\ee
where $24\pi^3$ is the 5D loop factor, and $g_4^2$ is computed, say, at the compactification scale $1/R$.
A closer inspection of the 5D vacuum polarization diagram shows that further factors of $\pi$ arise from the momentum integration. 
A more careful and conservative estimate would use the standard 4D loop factor to get
\be
\frac{g_5^2 \Lambda^{(2)}}{16\pi^2} = 1\ \ \ \Longrightarrow \Lambda^{(2)} = \frac{8\pi}{g_4^2} \frac{1}{R}\,,
\label{NDA2}
\ee
roughly a factor of 5 smaller than (\ref{NDA1}).
Another similar estimate can be given by comparing the one-loop term in eq.(\ref{RGg5d}) to the classical one, $g_5^{-2}(E_0)$. In this way, we get
\be
\Lambda^{(3)} = \frac{128}{3g_4^2}\frac{1}{R }\,.
\label{RGestimate}
\ee
All these estimates unambiguously show that there is not a parametrically large range in energies (when $g_4\sim 1$) where 5D theories
are calculable and reliable effective field theories.  In this situation a factor of a few in the cut-off estimate can make the difference in defining a model reliable or not and it is hence very important  to improve by any means in discerning between the above estimates or adding new ones.

\begin{figure}[t!]
\begin{minipage}[t]{0.465\linewidth} 
\begin{center}
\includegraphics*[width=\textwidth]{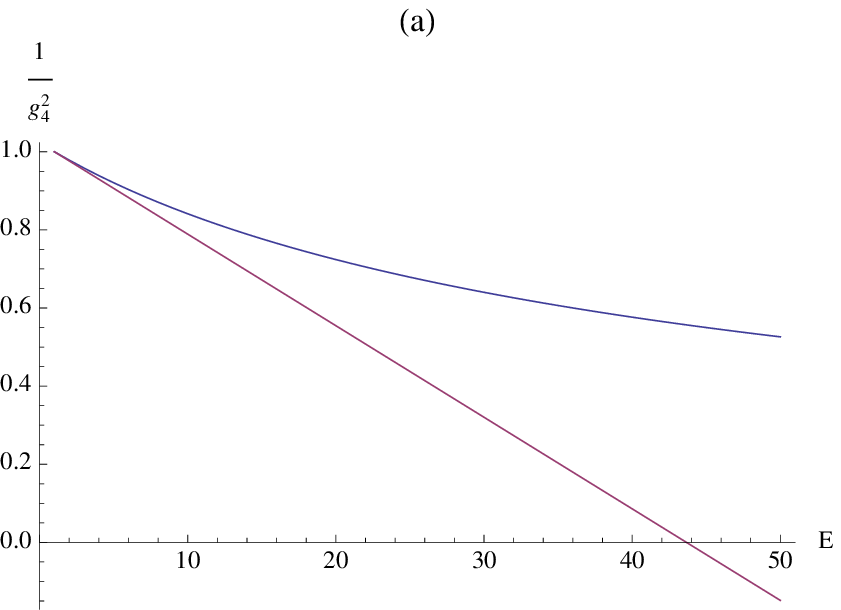}
\end{center}
\end{minipage}
\hspace{0.5cm} 
\begin{minipage}[t]{0.48\linewidth}
\begin{center}
\includegraphics*[width=\textwidth]{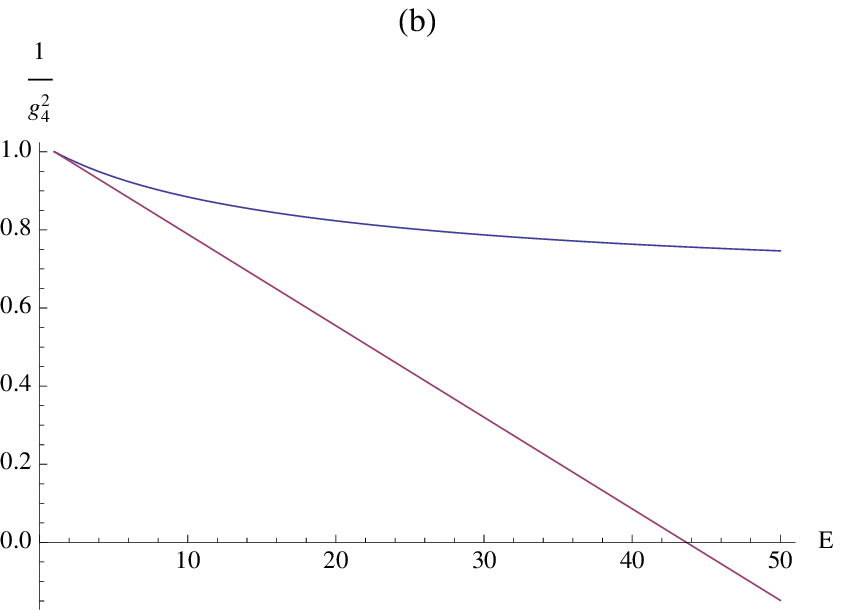}
\end{center}
\end{minipage}
\caption{Comparison between $g_4^{-2}$ as a function of the energy scale $E$ in the Lifshitz theory (blue line), as given by eq.(\ref{AnafE}),  and the effective 5D coupling (\ref{RGg5d}) (red straight line). 
The energy is in units of $1/R$. We have taken $E_0=1/R$ and  $g_4^{-2}(1/R)=1$. (a)  $a_\psi=1$ and $\Lambda = 25/R$.
(b) $a_\psi=10$ and $\Lambda = 100/R$. Despite the higher cut-off, the range of validity of the 5D effective theory is lower in (b) than in (a).} 
\label{figLifg}
\end{figure}

Our Lifshitz-like UV completion can be quite useful in this sense. 
 The cut-off scale $\Lambda$ should be here identified with the Lifshitz
cut-off scale (so far set to one)  which we now make explicit  as in section 2  by  setting $a_\psi\to a_\psi/\Lambda$ in our previous formulae.
The impossibility of having a too large window between $1/R$ and $\Lambda$ is clearly visible in 
our theory. If $1/R$ is too small  with respect to $\Lambda$, the 5D linear regime (\ref{RGg5d})
of $g^{-2}$ is too long before the Lifshitz operators comes to the rescue and perturbation theory breaks down.
We can compute which is the maximum allowed value for $\Lambda R$ by demanding that $g_4^{-2}(E)$ is definite positive for arbitrarily high energy scales.
Since for $E\gg \Lambda$, $f_4\simeq 2\pi R f_5$ goes to zero (see eq.(\ref{f5EUV})),
taking $E=\infty$ and $E_0=1/R$ in eq.(\ref{g25dE}), we get
\be
0\leq g_{4,\infty}^{-2} = g_4^{-2}-  f_{4}\Big(\frac{1}{\Lambda R},\frac{1}{\Lambda R}\Big) \,,
\label{g4Lambda}
\ee
where $g_4= g_4(E_0=1/R)$ and $g_{4,\infty}=g_4(E=\infty)$. Inverting eq.(\ref{g4Lambda}) to get
$\Lambda$ as a function of $R$ and $g_4$ is complicated. However, for $\Lambda R\gg 1$, 
${\cal O}(\Lambda R)$ KK modes significantly contribute to $f_4$ and
we  can safely replace $f_4$ by its non-compact version $f_5$, 
$f_4(1/\Lambda R,1/\Lambda R) \simeq 2\pi R f_5(1/\Lambda R)$. We can still approximate 
this expression by evaluating $f_{5}$ at zero:\footnote{We have numerically verified that the above two approximations lead to less than $10\%$ deviations 
 from the exact value for $\Lambda R\sim 10-30$.}
\be
f_{5}(0) = \frac{5 \Lambda}{96 a_\psi \pi^2} \,,
\ee
from which we finally find  (writing here $\Lambda\to\Lambda^{Lif.}$ to emphasize that it is the cut-off of the Lifshitz completion of the theory)
\be
\Lambda^{Lif.}  \gtrsim  \frac{48\pi a_\psi}{5}\frac{1}{R}\Big(\frac{1}{g_4^2}-\frac{1}{g_{4,\infty}^{2}}\Big) \,,
\label{LambdaLif}
\ee
where $a_\psi = a_\psi(\Lambda)$, essentially constant below $\Lambda$ \cite{Iengo:2009ix}.
It is reasonable that $\Lambda^{Lif.}$ depends on $a_\psi$, since the effective scale regulating the low-energy fermion propagators
is $\Lambda/a_\psi$ and not $\Lambda$. Increasing $a_\psi$, however, implies that the Lifshitz regime takes over the usual 5D regime earlier, and before
reaching the scale $\Lambda$ the 5D theory receive sizable UV-sensitive corrections.  This is illustrated in fig.5, where we show how increasing
$a_\psi$ allows a higher cut-off $\Lambda$, but does not change the range of validity of the effective field theory, which is always given by the scale 
(\ref{LambdaLif}) with $a_\psi=1$.

Comparing eq.(\ref{LambdaLif}) with eqs.(\ref{NDA1}), (\ref{NDA2}) and (\ref{RGestimate}), we see that for $a_\psi = 1$ the Lifshitz cut-off computation is closer to the most conservative of the three NDA cut-off estimates, eq.(\ref{NDA2}), provided that $g_{4,\infty}$ is not too small.
The more perturbative the UV-completion is, the lower the cut-off becomes. 

\section{IR Evolution of $c_\gamma$ and $c_\psi$, Fine-tuning and Astrophysical Constraints}

The recovery of Lorentz invariance at low-energy from the Lifshitz 5D QED is not automatic, since there is no mechanism enforcing
$c_\gamma=c_\psi$. These parameters evolve in the UV according to eq.(\ref{cevolution}), but their IR evolution is radically different.
For simplicity, we focus only on the 5D IR regime, $1/R<E<\Lambda$.
As explained before, in this regime it makes sense only a perturbative expansion in the coupling and thus the RG technique is not very useful.
 We will nevertheless continue to use this language for convenience and for homogeneity with the deep IR and UV regimes, in which the RG flow is useful.
 
When $E\ll \Lambda$, we can safely neglect the higher-derivative Lifhistz operators in the Lagrangian (\ref{Lquad}), and we end up with the usual 5D QED,
with $c_\gamma\neq c_\psi$. When $c_\gamma\neq c_\psi$,  Lorentz invariance is broken and both parameters run. The IR $\beta$-functions $\beta_{c_\psi}^{IR}$ and $\beta_{c_\gamma}^{IR}$ are easily determined. We take $c_\gamma = 1$ as initial condition, define $\delta c = c_\psi-c_\gamma$ and assume $\delta c\ll 1$, so that  we keep only up to linear terms in $\delta c$.  Let us first consider $\beta_{c_\gamma}^{IR}$, that can be determined by looking at the spatial components $\Pi_{mn}$ of the photon propagator corrections.
In the one-loop vacuum polarization photon diagram only one particle, the fermion, enters. Modulo a rescaling the graph is Lorentz invariant 
and correspondingly $\beta_{c_\gamma}^{IR}$ is proportional to the gauge coupling $\beta$-function. At linear order in $\delta c$, we find
\be
\beta_{c_\gamma}^{IR} = k_\gamma  g_5^2 E \delta c,
\label{betacgamma}
\ee
where $k_\gamma =- 3/(256\pi)$. 
The $\beta$-function $\beta_{c_\psi}^{IR}$  is determined by looking at the one-loop fermion propagator correction $\Sigma$. We define the functions
$f^{0,1}_\psi$ as
\be
\Sigma = i \gamma^0 \omega f^0_\psi  - i c_\psi \gamma_m p_m f^1_\psi\,\,.
\ee
In this way,
\be
\beta_{c_\psi}^{IR} = c_\psi E \frac{\partial}{\partial E} \Big(f_\psi^1-f_\psi^0\Big)\,.
\ee
Although the functions $f^{0,1}_\psi$ are gauge-dependent, the latter cancels in the difference so that  $\beta_{c_\psi}$
is gauge-invariant.  We get
\be
\beta_{c_\psi}^{IR} =k_\psi  g_5^2 E \delta c\,,
\label{betacpsi}
\ee
where $ k_\psi = 25/(1024 \pi)$.
Eqs.(\ref{betacgamma}) and (\ref{betacpsi}) are easily integrated giving
\bea
\delta c(E) \a = \a \delta c(E_0)  \big[1+g_5^2(E_0) k_\gamma (E-E_0)\big]^{\frac{k_\psi-k_\gamma}{k_\gamma}}\,,\nn \\
 c_\psi(E) \a = \a c_\psi(E_0)+ \delta c(E_0)\frac{k_\psi}{k_\psi-k_\gamma}  \big[1+g_5^2(E_0) k_g (E-E_0)\big]^{\frac{k_\psi-k_\gamma}{k_\gamma}}\,.\label{deltacSol}
\eea
Plugging the values found for $k_\gamma$ and $k_\psi$ in eq.(\ref{deltacSol}), we get
\be
\delta c(E) \simeq \frac{\delta c(E_0)}{[1+g_5^2(E_0) k_\gamma (E-E_0)]^3}\,.
\label{cEevolution}
\ee
The factor $\delta c(E)$ decreases towards the IR, as desired. 
Unfortunately, the decrease one gets is not very efficient to avoid the need of fine-tuning.
Even if we pretend that eq.(\ref{cEevolution}) is reliable beyond one-loop level, and take, for example, $E=1/R$, $E_0\simeq 25/R$, $g_4^2(E_0) \simeq 4\pi$ (strong coupling),
$\delta c$ decreases along the flow by
at most three orders of magnitude, whereas experimental bounds require $\delta c\lesssim 10^{-21}$ for ordinary particles \cite{Coleman:1998ti}
 (for electrons and photons the bounds are slightly less severe, see e.g. \cite{Kostelecky:2008ts}).
Nevertheless,  by appropriately tuning $\delta c(E_0)$, Lorentz invariance can always be achieved with the desired accuracy. 
In the deep IR regime, the evolution of $c_\psi$ and $c_\gamma$ will change from a linear to a logarithmic behaviour below $1/R$.
No new qualitative features emerge and we will not report the corresponding results.

Even if $c_\gamma=c_\psi=1$ to a sufficient precision in the deep IR, the dispersion relations of photon and electrons in our theory are modified:
\be
v_i(p)= \frac{d\omega_i}{dp}=\frac{2 a_i^2 p^3+c_i^2 p}{\sqrt{a_i^2 p^4+c_i^2 p^2}}  = 1+\frac{3a_i^2 p^2}{2}+{\cal O}(p^{-4})   \,, \ \ \ i=\psi, \gamma\,,
\label{vidisp}
\ee
leading to an energy-dependent maximum allowed speed for the two particles.  Astrophysical bounds, particularly coming from cosmic ray observations,
generally constraint the size of the $a_i p^2$ corrections above, pushing $\Lambda$ to very high scales 
for $a_i\sim 1$ (see e.g. \cite{Liberati:2009pf} for an overview). We have not systematically studied the bounds on $\Lambda$ coming from
these experiments, but considered only a specific one, which is quite likely not the most stringent one.  
It arises from the time delay measured by the  FERMI experiment in the gamma ray burst GRB 080916C at red-shift $z=4.35$ \cite{Abdo:2009zz}
and has the advantage of being purely kinematical.
This bound can be roughly cast in the following way
\be 
|v_\gamma^2({\rm 1MeV}) - v_\gamma^2(10{\rm GeV})| \lesssim 10^{-17}\,.
\label{cexp}
\ee
Reinserting the scale $\Lambda$ in eq.(\ref{vidisp}), the bound (\ref{cexp}) gives
\be
\Lambda \gtrsim 5\times 10^9 \; {\rm GeV}\,  a_\gamma \,.
\label{boundLambda}
\ee

\section{Conclusions}

We have constructed a renormalizable, UV completed, Lifshitz-like theory that reduces at low energies to the standard QED in 5D.
This is the simplest and most  concrete UV completion of a ED theory we are aware of, with excellent UV properties. In particular, the gauge coupling constant is finite to all orders in perturbation theory. The one-loop behaviour of the coupling is described, at {\it all} energy scales, by eq.(\ref{f4D}).
 We have shown in detail how eq.(\ref{f4D}) reproduces, as it should, the energy behaviour
of a coupling constant in 4 and 5 dimensions at lower energies. We have then derived a bound on the size of the cut-off in the 5D  QED theory,
based on our UV completion. Our results show that the often used NDA estimate (\ref{NDA1}) is too optimistic, while the more conservative estimate (\ref{NDA2})
is more reliable.

Admittedly, our UV completion is not very well motivated. One has to impose a severe fine-tuning to recover Lorentz invariance at 
low energies \cite{Iengo:2009ix,Collins:2004bp}.  Moreover, the Lifshitz cut-off is severely constrained by astrophysical data, as 
shown e.g. in eq.(\ref{boundLambda}). Nevertheless, we think our model can be useful, at least seen as a toy UV-completion mechanism of effective ED theories.
Several issues related to the calculability of higher dimensional non-renormalizable theories and the UV sensitivity of observables can concretely be 
addressed using generalizations of our QED Lifshitz construction. Given the simplicity of the theory, we think that all the necessary generalizations needed
to construct a UV completion of phenomenologically interesting models (interval compactifications, localized brane terms, warp factors, etc.) should not
represent a too complicated task.

\section*{Acknowledgments}

We would like to thank Stefano Liberati, Enrico Trincherini and especially Riccardo Rattazzi for useful discussions.

\appendix

\section{An Approximate Analytic Expression for $f_{5}(E)$}

This approximation is  found by decomposing the integral over $s$ appearing in eq.(\ref{f5D}) in two:
$\int_0^\infty ds = \int_0^1 ds+\int_1^\infty ds$, and simplifying the integrand in the two regimes as follows:
\be
f_{5}(E) \simeq \frac{1}{16\pi^2 a_\psi} \bigg[\int_0^1\!ds \frac{3s^2}{(4s^2+4s^4 +\mu^2)}+\int_1^\infty\!ds \frac{4s}{(4s^2+4s^4 +\mu^2)}\bigg] \,.
\label{appf5d}
\ee
The $s$ integration can now be performed and we obtain the following expression for $f_{5}(E)$:
\bea
f_{5}(E) \a = \a\frac{1}{128\pi^2a_\psi}\Bigg\{ \frac{3\sqrt{2}\Big[(-1+\sqrt{1-\mu^2})\arctan\left(\frac{\sqrt{2}}{\sqrt{1-\sqrt{1-\mu^2}}}\right)+
\mu\arctan\left(\frac{\sqrt{2}}{\sqrt{1+\sqrt{1-\mu^2}}}\right)\Big]}{\sqrt{1-\mu^2}\sqrt{1-\sqrt{1-\mu^2}}}\nn \\
\a\a \hspace{-1cm} -\frac{4\Big[\Big(\pi+2\arctan \left(\frac{3}{\sqrt{-1+\mu^2}}\right)\Big)\theta(1-\mu)
-\Big(\pi-2\arctan \left(\frac{3}{\sqrt{-1+\mu^2}}\right)\Big)\theta(\mu-1)\Big]}{\sqrt{\mu^2-1}} \Bigg\} ,
\label{AnafE}
\eea
with $\mu$ as in eq.(\ref{musDef}). Despite the appearance of negative square roots for any $\mu$, the
function $f_{5}(E)$ is real. Eq.(\ref{AnafE}) turns out to be a very good approximation of eq.(\ref{f5D}).

\end{document}